\documentclass[12pt]{article}

\usepackage[mathscr]{eucal}
\usepackage{amsfonts}
\usepackage{amssymb}
\usepackage{amsthm}

\newcommand{\be}{\begin{equation}}
\newcommand{\ee}{\end{equation}}
\newcommand{\bea}{\begin{eqnarray}}
\newcommand{\eea}{\end{eqnarray}}

\newcommand{\p}[1]{(\ref{#1})}

\begin{document}

\begin{titlepage}
\vspace*{1.5cm}

\renewcommand{\thefootnote}{\dag}
\begin{center}
{\LARGE\bf  Galilean  Conformal Mechanics}

\vspace{0.5cm}

{\LARGE\bf from Nonlinear Realizations}

\vspace{1.5cm}
\renewcommand{\thefootnote}{\star}

{\large\bf Sergey Fedoruk}${}^{\,1}$,\,\,\, {\large\bf Evgeny Ivanov}${}^{\,1}$,\,\,\,
{\large\bf Jerzy Lukierski}${}^{\,2}$ \vspace{1cm}

${}^{1)}${\it Bogoliubov  Laboratory of Theoretical Physics, JINR,}\\
{\it 141980 Dubna, Moscow region, Russia} \\
\vspace{0.1cm}

{\tt fedoruk@theor.jinr.ru, eivanov@theor.jinr.ru}\\
\vspace{0.8cm}

${}^{2)}${\it Institute for Theoretical Physics, University of Wroc{\l}aw,}\\
{\it pl.
Maxa Borna 9, 50-204 Wroc{\l}aw, Poland} \\
\vspace{0.1cm}

{\tt lukier@ift.uni.wroc.pl}\\
\vspace{0.4cm} \setcounter{footnote}{0}

\end{center}
\vspace{0.2cm} \vskip 0.6truecm  \nopagebreak

\begin{abstract}
\noindent We apply the nonlinear realizations method for constructing new Galilean conformal mechanics
models. Our starting point is the Galilean conformal algebra which is a non\-relativistic
contraction of its relativistic counterpart. We calculate
Maurer--Cartan one--forms, examine various choices of the relevant coset
spaces and consider the geometric inverse Higgs--type constraints which reduce the
number of the independent coset parameters and, in some cases,
provide dynamical equations.
New Galilean conformally invariant actions are derived in arbitrary
space--time dimension $D\,{=}\,d{+}1$ (no central charges), as well as in the
special dimension $D\,{=}\,2{+}1$ with one ``exotic'' central charge. We obtain new
classical mechanics models which extend the standard ($D\,{=}\,0{+}1$) conformal
mechanics in the presence of $d$ non-vanishing space dimensions.
\end{abstract}

\bigskip\bigskip\bigskip\bigskip
\noindent PACS: 11.10-z; 11.10.Kk; 11.25.Hf

\smallskip
\noindent Keywords: nonlinear realization, Galilei conformal group, inverse Higgs effect

\newpage

\end{titlepage}

\setcounter{equation}{0}
\section{Introduction}

In recent years, there was an increasing interest in the applications
of renowned AdS/CFT correspondence \cite{Mal,GuKlPol,Wit} to non--relativistic conformal
field theory \cite{Son,BalMcC} (see also \cite{BagGop,Ali,MarTach,Bagc} and references therein).
In the AdS/CFT framework, an important role is played by the
(super)conformal quantum mechanics as the simplest
counterpart of the higher-dimensional (super)conformal field
theories. Keeping in mind an extension to the non-relativistic case,
it is desirable to consider various non-relativistic versions of the (super)conformal mechanics. The
study of such models would allow to gain deeper insights into the
physical and mathematical aspects of non-relativistic conformal
symmetry and can be used in the analysis of the corresponding (super)strings and field theories.

The basic aim of the present paper is to construct, at the classical level, several new
mechanical models invariant under Galilean conformal symmetries. Our main tool will be
the systematic use of the universal geometric method of nonlinear realizations \cite{NonLin}.

It is known that the de Alfaro, Fubini, Furlan (AFF)  conformal
mechanics \cite{AFF}, as well as its supersymmetric extensions
\cite{AP,FR,IKL2}, can be adequately described in the framework of
nonlinear realizations of the $D{=}\,0{+}1$ conformal group
SL(2,R)$\sim $SO(1,2)~\cite{IKL} and its supersymmetric extensions
\cite{IKL2,AIPT}. An important part of these geometric techniques is
the covariant reduction of the number of (super)conformal group
parameters by means of the inverse Higgs mechanism \cite{IO}, which
singles out the dynamical variables. The inverse Higgs constraints
can be derived in the geometrically transparent way, using the
formalism of the Maurer-Cartan (MC) one--forms on the suitably
chosen cosets of the symmetry group.

In this paper we apply the MC method to the Galilean conformal (GC)
group. The GC group extends the one--dimensional conformal symmetry
of \cite{AFF} to the conformal symmetry of the $D{=}\,d{+}1$ -- dimensional non-relativistic
space--time, with $d \geq 1$ being the number of space dimensions. For a
long time, since it was proposed in \cite{Ha,Ni}, the name of
non-relativistic conformal symmetry was attributed to the
Schr\"{o}dinger symmetries, which provide the covariance of the
Schr\"{o}dinger equation describing non-relativistic massive
particle \footnote{It should be mentioned that Schr\"{o}dinger
algebra is simply related with the Lie algebra description of
invariance of heat equation, proposed firstly in XIX-th century
\cite{Lie}.}. However, the corresponding Schr\"{o}dinger algebra does not require
mass parameters to vanish and does not contain the non-relativistic
counterpart of the conformal spatial accelerations. An alternative
candidate for the non-relativistic conformal symmetry algebra is the
Galilean conformal algebra (GCA), and it is the symmetry we shall deal with in this paper.
It can be obtained by a
contraction of the $\frac{(d+2)(d+3)}{2}$ - dimensional relativistic
conformal algebra $o(d+1,2)$, in such a way that the number of
generators
is preserved \cite{Barut,HavPl,Henkel,NOR,LSZ2,DuHo} \footnote{To avoid a possible confusion, let us note
that the term ``non-relativistic conformal symmetries'' is sometimes used
for the infinite--dimensional conformal isometries of
non-relativistic space--time \cite{HavPl,Duval,DuHo,BagGop}.
They arise due to the degeneracy of the Galilean space-time
metric and have no relativistic counterpart. Here we will not deal with
this type of conformal symmetries.}.

GCA in $d$ space dimensions has the following semi-direct sum structure
\begin{equation}
\label{GCA-Str}
{\mathscr{C}}^{(d)}=\Big( o(2,1)\oplus o(d) \Big) \ltimes {\mathscr{A}}^{(3d)}\,,\;(d\geq 2)\,, \quad
{\mathscr{C}}^{(1)}= o(2,1)\ltimes {\mathscr{A}}^{(3)}\,, \quad {\mathscr{C}}^{(0)} = o(2,1)\,.
\end{equation}
Here, $o(2,1)$ describes the conformal symmetries on the world line and was employed in \cite{AFF,IKL},
$o(d)$ generates the space rotations and ${\mathscr{A}}^{(3d)}$ represents the $3d$--dimensional Abelian subalgebra
of space translations, Galilean boosts and non-relativistic constant accelerations.
We see that the symmetry of standard conformal mechanics is given by ${\mathscr{C}}^{(0)}= o(2,1)\subset {\mathscr{C}}^{(d)}$
for $d{>}0$;
besides, it is clear that  ${\mathscr{C}}^{(d)}$ includes as a subalgebra the centerless
Galilean $\frac{(d+1)(d+2)}{2}$ -- dimensional algebra
in $d$ space dimensions:
\begin{equation}\label{GA-Str}
{\mathscr{G}}^{(d)}=\Big( o(1,1)\oplus o(d) \Big) \ltimes \tilde{\mathscr{A}}^{(2d)}\,,\;(d\geq 2)\,, \quad
{\mathscr{G}}^{(1)}= o(1,1)\ltimes \tilde{\mathscr{A}}^{(2)}\,,
\end{equation}
where the Abelian subalgebra $o(1,1)$ describes the time translations and
$\tilde{\mathscr{A}}^{(2d)}$ is formed by the space translations and Galilean boosts, i.e. we get that
${\mathscr{G}}^{(d)}\subset {\mathscr{C}}^{(d)}\,$. Note that the algebra \p{GA-Str}  admits an extension by one central charge
for any $d$ and by two central charges for the special case of $d\,{=}\,2$ (see below); however, only the latter case, and with only
one central charge, can be promoted to a central extension of the GCA for $d{=}\,2$.
We also add that the semi-direct sum structure represented by the formulae  (\ref{GCA-Str}), (\ref{GA-Str})
reflects as well the semi-direct product decomposition of the corresponding centerless Galilean and Galilean conformal groups.
We will denote these groups as $\hat{\mathscr{G}}^{(d)}$ and $\hat{\mathscr{C}}^{(d)}$, respectively.

We begin our paper by recalling, in Sect.\,2, a few known examples of the application of the techniques of
MC one--forms and inverse Higgs constraints to mechanical systems in order to derive
the relevant dynamical equations of motion and invariant actions. We consider the following models of classical mechanics:

\noindent ${\bf i.}$ \,\, Massive free non-relativistic particle model \\
$\phantom{1}$ \,\,\, (for any $d$, we use ${\mathscr{G}}^{(d)}$ with
one central charge $M$) \cite{AIM-F,GGT};

\noindent ${\bf ii.}$ \,Massive free non-relativistic particle model with higher order
Chern--Simons--type term  \\
$\phantom{1}$ \,\,\, (for $d{=}2$, we use ${\mathscr{G}}^{(2)}$ with
two central charges) \cite{LSZ};

\noindent ${\bf iii.}$ Standard conformal mechanics model (for $d{=}0$, with ${\mathscr{C}}^{(0)}=o(2,1)$) \cite{AFF,IKL}.

In cases ${\bf i.}$, ${\bf ii.}$ we shall use the MC one--forms for centrally extended Lie algebras.
In particular, the MC one--forms associated with the central charge generators will be used for the
geometric construction of the invariant actions \cite{AIM-F,GGT,AGKP}.

In Sect.\,3 we consider GCA for arbitrary $d$, calculate the corresponding MC one--forms and propose
the choice of inverse Higgs constraints and GC--covariant dynamical equations,
which leads to the extensions of standard AFF conformal mechanics model \cite{AFF,IKL}.
We consider four examples of cosets for GCA. The canonical choice
(with the stability subalgebra $o(d)$) is shown to lead, after imposing the properly chosen inverse Higgs constraints,
to new  GC--covariant field equations for arbitrary $d$.

In Sect.\,4 we propose the actions dynamically generating the GC--covariant inverse Higgs constraints.
For arbitrary $D{=}\,d{+}1$ we propose two new extensions of the AFF
model \cite{AFF} and, for $D\,{=}\,2{+}1$, an extension of the conformal dynamics considered in \cite{SZ}.

Brief conclusions are collected in Sect.\, 5.

\setcounter{equation}{0}
\section{Nonlinear realizations method in classical mechanics: illustrative examples}

In this section we recall some known examples of the nonlinear realizations:
for the standard Galilei and exotic Galilei groups and for one--dimensional conformal group ${\mathscr{C}}^{(0)}$
leading to standard conformal mechanics.
We demonstrate that the dynamics of considered systems is completely determined by imposing the appropriate
conditions on the MC one--forms. Some of these conditions, the inverse Higgs constraints,
are algebraic equations eliminating part of the original coset variables.
Other covariant conditions imposed on the MC one--forms are the dynamical equations of motion.
Invariant actions are also constructed by making use of the MC one--forms.
In nonconformal cases the correct
action is obtained from the MC one--forms associated with the generators of central charges.
In the case of conformal mechanics,
both the algebraic constraints and the equations of motion follow from the action which is linear in MC one--forms.

\subsection{Galilei group in arbitrary space--time dimension $D$}

The centrally extended Galilei algebra in $D{=}\,d{+}1$--dimensional space--time is spanned by
the generator of the time translation $H$, the space translation generators $P_i$, $i=1,...,d$,
the boosts $B_i$, the $o(d)$ rotation generators $J_{ij}=-J_{ji}$ and the central charge $M$ describing a non-relativistic mass.
The full set of commutation relations consists of $o(d)$ Lie algebra, the relations
\begin{equation}\label{Gal-al1}
[H,P_{i}]=0 \,,\qquad [H,B_{i}]=iP_{i} \,,
\end{equation}
\begin{equation}\label{Gal-al2}
[P_{i},P_{j}]=0 \,,\qquad [B_{i},P_{j}]=i\delta_{ij}M \,,\qquad [B_{i},B_{j}]=0
\end{equation}
and the commutators of $o(d)$ generators $J_{ij}$ with the vector generators  $P_i$, $B_i$.

Let us consider the nonlinear realization of
the centrally extended Galilei group in the coset with O$(d)$ as the stability subgroup \cite{AIM-F,GGT}. We choose the following
explicit parametrization of the coset
\begin{equation}\label{group-el-G}
G = e^{itH}\, e^{ix_kP_k}\, e^{iv_kB_k}\, e^{i\varphi M}\,.
\end{equation}
The left--invariant MC one--forms defined by the general relation
\begin{equation}\label{om-def-G}
G^{-1}dG = i\Big(\omega_H H+
\omega_{P\!,\,k} P_k + \omega_{B\!,\,k} B_k  + \omega_{M} M\Big)
\end{equation}
are given by the following explicit expressions
\begin{equation}\label{om-G}
\omega_H = dt\,,\qquad
\omega_{P\!,\,i} = dx_i - v_i\,dt\,,\qquad \omega_{B\!,\,i} = dv_i\,,\qquad
\omega_M = d\varphi+v_idx_i-{\textstyle\frac12}\,v_i v_i \,dt\,.
\end{equation}

The group variables in (\ref{group-el-G}) describe the mechanical system with the Hamiltonian $H$ and the trajectories
in the extended phase space $x_i=x_i(t)$, $v_i=v_i(t)$, $\varphi=\varphi(t)$.
The fields $v_i(t)$ can be covariantly eliminated by imposing the algebraic inverse Higgs constraints
\begin{equation}\label{inH-G}
\omega_{P\!,\,i} =0 \qquad \Rightarrow \qquad v_i=\dot x_i\,.
\end{equation}
The equations of motion for the remaining physical variables $x_i(t)$ are represented by the  constraints
\begin{equation}\label{dynur-G}
\omega_{B\!,\,i} =0 \qquad \Rightarrow \qquad \ddot x=0 \,.
\end{equation}
Both the algebraic inverse Higgs constraints  (\ref{inH-G}) and dynamical equations (\ref{dynur-G})
can be derived as the Euler--Lagrange equations from the first-order action \cite{AIM-F,GGT}
\begin{equation}\label{ac-G}
S_m=m\!\int \omega_{M} = m\!\int dt\left[\dot\varphi+v_i\dot x_i -{\textstyle\frac12}\,v_i v_i \right].
\end{equation}
After inserting the constraints  (\ref{inH-G}) in the action (\ref{ac-G})
we obtain (up to a total derivative under the integral) the standard action for the massive particle
\begin{equation}\label{ac-G-c}
S_m={\textstyle\frac{1}{2}}\,m\!\int dt\,\dot x_i \dot x_i .
\end{equation}

\subsection{Exotic Galilei group in $D\,{=}\,2{+}1$}

The $D{=}2{+}1$--dimensional space--time is special because in this case we can add
to the set $H$, $P_i$, $B_i$, $J_{ij}$, $M\,$
the second (exotic) central charge $\Theta$ and consider exotic Galilei algebra
with the additional non-vanishing commutators
\begin{equation}\label{eGal-al}
[B_{i},B_{j}]=i\epsilon_{ij}\Theta \,.
\end{equation}
The MC one--forms
\begin{equation}\label{om-def-eG}
\tilde G^{-1}d\tilde G = i\Big(\omega_H H+
\omega_{P\!,\,k} P_k + \omega_{B\!,\,k} B_k  + \omega_{M} M + \omega_{\Theta} \Theta\Big)\,,\qquad
\tilde G = G\, e^{i\phi\, \Theta}\,,
\end{equation}
are given by (\ref{om-G}) and the additional expression
\begin{equation}\label{om-eG}
\omega_\Theta = d\phi+{\textstyle\frac12}\,\epsilon_{ij}v_idv_j
\end{equation}
for the MC one--form corresponding to the exotic central charge.
Using this one--form, we can generalize (\ref{ac-G}) and
consider the action \cite{AGKP}
\begin{equation}\label{ac-eG}
S_{m,\theta}=m\!\int \omega_{M}+ \theta\!\int \omega_{\Theta}= m\!\int dt
\left[\dot\varphi+v_i\dot x_i -{\textstyle\frac12}\,v_i v_i\right]
+{\textstyle\frac{1}{2}}\,\theta \! \int dt\left[2\dot\phi+\epsilon_{ij}v_i\dot v_j \right],
\end{equation}
where both $\theta$ and $m$ are constant. Inserting the inverse Higgs constraints
(\ref{inH-G})\footnote{If $\theta\neq 0$ the constraints  (\ref{inH-G}) can be obtained from (\ref{ac-eG})
as on--shell conditions, i.e. as a consequence of field equations.},
we get, modulo a total derivative, the action
for $D{=}2{+}1$ massive particle with the higher-order
Chern--Simons--type term, which was proposed in \cite{LSZ}:
\begin{equation}\label{ac-eG-c}
S_{m,\theta}= {\textstyle\frac{1}{2}}\,m\!\int dt \,\dot x_i\dot x_i
+{\textstyle\frac{1}{2}}\,\theta\!\int dt\,\epsilon_{ij}\dot x_i \ddot{x}_j \,.
\end{equation}

\subsection{$D\,{=}\,0{+}1$ conformal mechanics}

Following \cite{IKL}, the AFF conformal mechanics \cite{AFF}
can be obtained by applying the MC method to the one--dimensional conformal algebra ${\mathscr{C}}^{(0)}= o (2,1)$:
\begin{equation}\label{SL2R}
[D,H] = - iH \, , \qquad [K,H]=-2iD \, , \qquad [D,K]=iK\,.
\end{equation}

We choose the exponential parametrization for the  group
$\hat{\mathscr{C}}^{(0)}{=}\,{\rm SO}(2,1)$\,:
\begin{equation}\label{group-el-cm}
\hat{\mathscr{C}}^{(0)}\equiv G_0 = e^{itH}\, e^{izK}\, e^{iuD}\,,
\end{equation}
and obtain the following left--covariant MC one--forms
\begin{equation}\label{om-def0}
G_0^{-1}dG_0 = i\Big(\omega_H H+ \omega_K K+ \omega_D D \Big)\,,
\end{equation}
with
\begin{equation}\label{om-cm}
\omega_H = e^{-u}dt\,,\qquad
\omega_K = e^{u}\left(dz+z^2 dt\right)\,,\qquad
\omega_D = du-2zdt\,.
\end{equation}

In the conformal mechanics model \cite{AFF}, as in the construction of unitary representations of the group SO(2,1) \cite{Barg},
one is led to the choice of the following basis in the $o(2,1)$ algebra
\begin{equation}\label{new-o21}
R^\pm={\textstyle\frac12}\left( \gamma K\pm \gamma^{-1}H\right)\,,\quad\,\, D\,;
\qquad\quad [R^+,R^-]=iD \, , \qquad [D,R^\pm] = iR^\mp \,,
\end{equation}
where $\gamma$ is a constant with the mass dimension, so that $R^\pm$ are dimensionless.
The MC one--forms related with the generators $R^\pm$ are, respectively,
\begin{equation}\label{new-o21-CM}
\omega_R^\pm= \gamma^{-1}\omega_K\pm \gamma \omega_H\,.
\end{equation}

The dynamics of AFF conformal mechanics is obtained by imposing the following
constraints~\cite{IKL}
\begin{equation}\label{IKL-const}
\mbox{(a)}\,\,\,\,\,\omega_D=0\,,\qquad \qquad \mbox{(b)}\,\,\,\,\,\omega^-_{R}=0
\end{equation}
on the one--dimensional coset fields $z(t)$ and $u(t)$. {}From the inverse Higgs constraint (\ref{IKL-const}a) it follows that
\begin{equation}\label{z-rez}
z={\textstyle\frac12}\,\dot u\,,
\end{equation}
while the dynamical constraint (\ref{IKL-const}b) leads to the equation of motion
\begin{equation}\label{eq-rho}
\ddot\rho=\gamma^2\rho^{-3}
\end{equation}
for the single independent variable $\rho=e^{u/2}$.

The standard AFF conformal mechanics action \cite{AFF}
\begin{equation}\label{ac-AFF}
S_0= \int dt \,\left(\,\dot \rho^2 -\gamma^2\rho^{-2}\,\right),
\end{equation}
which generates the equation of motion (\ref{eq-rho}), in the formalism of MC one--forms can be rewritten as~\cite{IKL}
\begin{equation}\label{ac-cm-f}
S_0=-\gamma\int \omega_{R}^+ = -\int\,dt\Big[ \,e^u\left(\dot z + z^2 \right) +\gamma^2 e^{-u} \,\Big]\,.
\end{equation}
We see that the action (\ref{ac-cm-f}) is specified by the remaining non-vanishing MC one--form in $o(2,1)$.
Both the kinematical constraint (\ref{z-rez})
($\omega_D=0$) and the dynamical
equation (\ref{eq-rho}) ($\omega^-_{R}=0$), are the equations of motion following from the action (\ref{ac-cm-f}).

Note that the equations (\ref{IKL-const}) define a class of geodesics on the ${\rm SO}(1,2)$ group manifold,
described by the one--parameter compact subgroup with the generator $R^+$ \cite{IKL}. Only such a class leads
to the standard conformal mechanics with good quantum properties \cite{AFF}, as opposed to any other non-trivial choice
of the constraints (for example, the choice of $\omega^+_{R}=0$ instead of (\ref{IKL-const}b)).
This is the reason why in our further considerations we will use only the constraints (\ref{IKL-const}).

Let us make brief comments on the Hamiltonian formulation of the model (\ref{ac-cm-f}), which will be useful later in the
consideration of other GC invariant actions.

The definitions of the momenta yield the second class constraints
\begin{equation} \label{constr-cm}
p_u  \approx 0\,, \qquad
p_z +e^u \approx  0\,.
\end{equation}
These constraints allow to eliminate the phase space variables $(p_z, p_u)\,$.
The Dirac brackets for the surviving pair of the phase space variables $(u,z)$ and the Hamiltonian take the form
\begin{equation}\label{tHam1-cm}
\{ u, z\}_{{}_D}=e^{-u}\,,\qquad H= e^u z^2  +\gamma^2 e^{-u}\,.
\end{equation}
After introducing the variables $\rho=e^{u/2}$, $p_\rho=2 e^{u/2}z$ which possess the standard canonical brackets
\begin{equation}\label{can-tr-cm}
(u, z):\,\,\,\{ u, z\}_{{}_D}=e^{-u} \qquad \Rightarrow  \qquad (\rho, p_\rho):\,\,\,\{ \rho, p_\rho\}_{{}_D}=1\,,
\end{equation}
we obtain that the system (\ref{ac-cm-f}) is described by the Hamiltonian
\begin{equation}\label{tHamAFF-cm}
H={\textstyle\frac{1}{4}}\,p_\rho^2+ \gamma^2 \rho^{-2}\,,
\end{equation}
which follows from the action (\ref{ac-AFF}).

\setcounter{equation}{0}
\section{Algebraic description of Galilean conformal symmetry}

\subsection{Galilean conformal algebra and corresponding MC one--forms}

\subsubsection{Arbitrary $D$}

Galilean conformal algebra (GCA) ${\mathscr{C}}^{(d)}$ in $D\,{=}\,d{+}1$
defined by eq.\,(\ref{GCA-Str}) is obtained by adding
to the ${\mathscr{C}}^{(0)}=o(2,1)$ algebra (\ref{SL2R})
the Lie algebra of space rotations $\emph{o}(d)$ which commutes  with ${\mathscr{C}}^{(0)}$:
\begin{equation}\label{M-M}
[J_{ij}, J_{k l}]= i\left(\delta_{ik} \,J_{j l} - \delta_{i l}\, J_{jk} + \delta_{j l}\,
J_{ik} - \delta_{jk} J_{i l}\right)\,,
\end{equation}
\begin{equation}\label{M-T}
[J_{ij}, H]=[J_{ij}, D]=[J_{ij}, K]=0\,,
\end{equation}
as well as the $3d$--dimensional Abelian subalgebra ${\mathscr{A}}^{(3d)}$ spanned by the generators
$P_i$, $B_i$ and $F_i$ with the following commutators
\begin{equation}\label{H-A}
\begin{array}{llll}
& [H,P_k]=0 \, , \qquad  &[H,F_k]=2iB_k\, , \qquad   &[H,B_k]= iP_k \, ,
\\
& [K, P_k]=-2iB_k \, , \qquad  &[K, F_k]=0 \, , \qquad  &[K,B_k]=-iF_k\, ,
\\
& [D,P_k]=- iP_k \, , \qquad   &[D,F_k]=iF_k \, , \qquad  &[D,B_k]=0 \, ,
\end{array}
\end{equation}
\begin{equation}\label{M-A}
[J_{ij},{\mathscr{A}}_{a,k}]=i\left(\delta_{ik} {\mathscr{A}}_{a,j} - \delta_{jk} {\mathscr{A}}_{a,i}\right) \,,
\end{equation}
\begin{equation}\label{A-A}
[{\mathscr{A}}_{a,i},{\mathscr{A}}_{b,j}]=0 \,.
\end{equation}
Here, ${\mathscr{A}}_{1,i}=P_i$, ${\mathscr{A}}_{2,i}=B_i$ and ${\mathscr{A}}_{3,i}=F_i$.

The generators of GCA enlarge the algebras (\ref{Gal-al1}), (\ref{Gal-al2}) and (\ref{SL2R}) considered in the previous section.
We recall that the operators $B_i$ generate the
Galilean boosts and the non-relativistic energy operator $H$ generates the Galilean time
translations. The operators $F_i$ generate constant non-relativistic accelerations,
and their presence implies that
the central charge $M$ (introduced in (\ref{Gal-al2})) should be put equal to zero for any $D =d +1$. For $d=2$ one can still add
the central charge $\Theta$ as in \p{eGal-al}, without breaking any Jacobi identity of the full GCA algebra ${\mathscr{C}}^{(2)}\,$.

We choose the coset ${\mathscr{K}}^{(d)}=\hat{\mathscr{C}}^{(d)}/{\mathscr{H}}$,
where $\hat{\mathscr{C}}^{(d)}$ is the GC group with  the algebra  (\ref{GCA-Str})
and ${\mathscr{H}}={\rm SO}(d)$. We call this coset {\it canonical} and use for it the following parametrization
\begin{equation}\label{group-el}
{\mathscr{K}}^{(d)} = G_0 \, e^{ix_kP_k}\, e^{if_kF_k}\, e^{iv_kB_k} \,,
\end{equation}
where $\hat{\mathscr{C}}^{(0)}=G_0$ is defined in (\ref{group-el-cm}).

The left--covariant MC one--forms are defined, as usual, by
\begin{equation}\label{om-def}
{\mathscr{K}}^{(d)}{}^{-1}d{\mathscr{K}}^{(d)} = i\Big(\omega_H H+ \omega_K K+ \omega_D D+
\omega_{P\!,\,k} P_k + \omega_{F\!,\,k} F_k + \omega_{B\!,\,k} B_k  \Big)\,.
\end{equation}
The forms $\omega_H$, $\omega_K$ and $\omega_D$ are the same as in (\ref{om-cm}), while
the remaining Cartan forms read
\begin{eqnarray}
\omega_{P\!,\,i} &=& dx_i + x_i\,\omega_D - v_i\,\omega_H\,,\label{om-P}\\
\omega_{F\!,\,i} &=& df_i - f_i\,\omega_D + v_i\,\omega_K\,,\label{om-F}\\
\omega_{B\!,\,i} &=& dv_i + 2\,x_i\,\omega_K - 2\,f_i\,\omega_H\,.\label{om-B}
\end{eqnarray}

Rewriting (\ref{om-def}) in the form
\begin{equation}\label{cov-der-def}
{\mathscr{K}}^{(d)}{}^{-1}d{\mathscr{K}}^{(d)} = i\omega_H \Big( H+ \mathscr{D}z\, K+ \mathscr{D}u \,D+
\mathscr{D}x_{k}\, P_k + \mathscr{D}f_{k}\, F_k + \mathscr{D}v_{k}\, B_k  \Big)
\end{equation}
we are left with the world-line density $E$:
\begin{equation}\label{wl-dens}
\omega_H = dt\,E\,, \quad E = e^{-u}\,,
\end{equation}
and the covariant time derivatives
\begin{equation}\label{D-all}
\begin{array}{rcl}
\mathscr{D}z &=& e^{2u}\left( \dot z+ z^2\right) \,, \\
\mathscr{D}u &=& e^{u} \left( \dot u -2 z \right)\,, \\
\mathscr{D}x_{i} &=& e^{u}\dot x_i + x_i\,\mathscr{D}u - v_i \,, \\
\mathscr{D}f_{i} &=& e^{u}\dot f_i - f_i\,\mathscr{D}u + v_i\,\mathscr{D}z\,, \\
\mathscr{D}v_{i} &=& e^{u}\dot v_i + 2\,x_i\,\mathscr{D}z - 2\,f_i \,.
\end{array}
\end{equation}

The infinitesimal transformations of the coset parameters, generated by the constant coset group elements
\begin{equation}\label{group-el-inf}
{\mathscr{K}}^{(d)}{}(\varepsilon) = e^{iaH}\, e^{ibK}\, e^{icD}\, e^{ia_kP_k}\, e^{ib_kF_k}\, e^{ic_kB_k}\,,
\end{equation}
are as follows
\footnote{We use the formula
$i{\mathscr{K}}^{(d)}{}^{-1}(\varepsilon\cdot T){\mathscr{K}}^{(d)}{}=
{\mathscr{K}}^{(d)}{}^{-1}\delta {\mathscr{K}}^{(d)}{} +\delta h$, where
$T$ are coset generators and
$\delta h$ defines induced transformations of the
stability subgroup $h_{ind}=1+\delta h$ (see \cite{Zum}).}
\begin{equation}\label{tr-GC}
\begin{array}{rcl}
\delta t &=& a+bt^2+ct\equiv \alpha(t) \,, \\
\delta z &=& b(1-2tz)-cz\,, \\
\delta u &=& 2bt+c\,, \\
\delta x_i &=& e^{-u}\left[a_i+t^2b_i+tc_i \right]\,, \\
\delta f_i &=& e^{u}\left[z^2a_i+(1-tz)^2b_i-z(1-tz)c_i \right]\,, \\
\delta v_i &=& -2za_i+2t(1-tz)b_i+(1-2tz)c_i \,.
\end{array}
\end{equation}
The forms (\ref{om-cm}) and (\ref{om-P})--(\ref{om-B}) are invariant with respect to the transformations (\ref{tr-GC})
and are covariant under the ${\rm SO}(d)$ transformations, which act as the standard  rotations of the vector index $i\,$.

In our further consideration, by analogy with the basis (\ref{new-o21}) in the $o(2,1)$ algebra,
we shall use the following new basis in the Abelian subalgebra ${\mathscr{A}}^{(3d)}$
\begin{equation}\label{new-Ab}
A^\pm_i= {\textstyle\frac12}\left( \gamma F_i\pm \gamma^{-1}P_i\right)
\,,\qquad B_i\,.
\end{equation}
The commutation relations between the generators  (\ref{new-Ab}) and (\ref{new-o21}) are as follows
\begin{equation}\label{R+-A}
\begin{array}{llll}
&[R^\pm,A^\pm_k]=0 \,,\quad   & [R^\pm,A^\mp_k]= \pm iB_k \,,\quad   &[R^\pm,B_k]=-iA^\mp_k\,,\\
&[D,A^\pm_k]= iA^\mp_k \,,\quad  & [D,B_k]=0 \,.\quad  &
\end{array}
\end{equation}
The explicit expressions for the corresponding MC one--forms
\begin{equation}\label{Cf-Ab}
\omega_{A\!,\,i}^\pm= \gamma^{-1} \omega_{F\!,\,i} \pm\gamma \omega_{P\!,\,i}
\,,\qquad \omega_{B\!,\,i}
\end{equation}
are
\begin{equation}\label{om-A}
\omega_{A\!,\,i}^\pm= d{\mathscr{X}}_{\,i}^{\,\pm} - {\mathscr{X}}_{\,i}^{\,\mp}\,\omega_D + v_i\,\omega_R^\mp\,,\qquad
\omega_{B\!,\,i} = dv_i + {\mathscr{X}}_{\,i}^{\,+}\,\omega_R^- - {\mathscr{X}}_{\,i}^{\,-}\,\omega_R^+\,,
\end{equation}
where we introduced new group variables
\begin{equation}\label{y-def}
{\mathscr{X}}_{\,i}^{\,\pm} = \pm\,  \gamma\, x_i + \gamma^{-1}f_i \,.
\end{equation}
The covariant derivatives of the new vector coset variables (\ref{y-def}) are
\begin{equation}\label{D-new}
\begin{array}{rcl}
\mathscr{D}{\mathscr{X}}_{\,i}^{\,\pm} &=& e^{u}\dot {\mathscr{X}}_{\,i}^{\,\pm} - {\mathscr{X}}_{\,i}^{\,\mp}\,\mathscr{D}u -
\gamma^{-1}v_i \left(\mathscr{D}z\mp \gamma^2  \right)\,, \\
\mathscr{D}v_{i} &=& e^{u}\dot v_i -\gamma^{-1}{\mathscr{X}}_{\,i}^{\,-} \left(\mathscr{D}z+ \gamma^2  \right)
+\gamma^{-1}{\mathscr{X}}_{\,i}^{\,+} \left(\mathscr{D}z- \gamma^2  \right) \,.
\end{array}
\end{equation}

\subsubsection{``Exotic'' $D{=}\,2{+}1$ case  with central charge $\Theta$}

If  $D\,{=}\,2{+}1\,$, the central charge $\Theta$ can be added (see (\ref{eGal-al})). It appears in the following commutators:
\begin{equation}\label{B-B-e}
[B_{i},B_{j}]=i\,\epsilon_{ij}\, \Theta \,,
\qquad
[P_{i},F_{j}]=-2i\,\epsilon_{ij}\, \Theta \,.
\end{equation}

The parametrization of the coset $\tilde{\mathscr{K}}^{(2)}=\tilde{\mathscr{C}}^{(2)}/{\rm{SO}}(2)$,
where $\tilde{\mathscr{C}}^{(2)}$ is the centrally extended GC group for $d{=}2$, can be chosen as
\begin{equation}\label{group-el-exo}
\tilde {\mathscr{K}}^{(2)} = G_0 \, e^{ix_kP_k}\, e^{if_kF_k}\, e^{iv_kB_k}\, e^{i\phi \Theta}\,,
\end{equation}
where $G_0=\hat{\mathscr{C}}^{(0)}$ and $k{=}1,2$.

The left--covariant MC one--forms are defined as
\begin{equation}\label{om-def-exo}
\tilde {\mathscr{K}}^{(2)}{}^{-1}d\tilde {\mathscr{K}}^{(2)} = i\Big(\omega_H H+ \omega_K K+ \omega_D D+
\omega_{P\!,\,k} P_k + \omega_{F\!,\,k} F_k + \omega_{B\!,\,k} B_k  + \omega_{\Theta} \Theta\Big)\,.
\end{equation}
All ``non-central'' one-forms are given by the old expressions (\ref{om-cm}) and (\ref{om-P})-(\ref{om-B}),
whereas $\omega_\Theta$ is
\begin{equation}\label{om-Th}
\omega_\Theta =d\phi-2\epsilon_{ij}f_i\,\omega_{P\!,\,j}+ {\textstyle\frac12}\,\epsilon_{ij\,}v_i\,\omega_{B\!,\,j}+
\epsilon_{ij\,}v_i \left(f_j\,\omega_{H} +x_j\,\omega_{K}\right)\,.
\end{equation}
The MC one--form $\omega_\Theta$ will be used in Sect.\,4 for the construction of new GC--invariant action.

To summarize, we observe that, before imposing the inverse Higgs constraints, our mechanical system
is spanned by the trajectories
\begin{equation}\label{coset-var}
z=z(t)\,,\qquad u=u(t)\,,\qquad x_k=x_k(t)\,,\qquad f_k=f_k(t)\,,\qquad v_k=v_k(t)\,,
\end{equation}
describing the motion in the coset ${\mathscr{K}}^{(d)}$, and, in the specific $D{=} 2{+} 1$ case,  in the coset $\tilde {\mathscr{K}}^{(2)}$
with the extra coordinate $\phi(t)\,$. In the next subsection, we propose the natural covariant constraints on the MC one--forms
which permit to eliminate a part of the functions (\ref{coset-var}).

\subsection{Inverse Higgs constraints and field equations}

In this subsection and in Section 4 we shall consider possible extensions of the AFF conformal mechanics which are covariant
under the Galilean conformal symmetry with $d{\neq}0$.
Here we start our analysis at the level of equations of motion, leaving aside the existence
of relevant Lagrangians.
The choice of independent dynamical degrees of freedom is specified by the choice of cosets and the appropriate
inverse Higgs constraints. The dynamical equations are also formulated as constraints imposed on the MC one-forms.
In such a way resulting dynamics is by construction covariant under the GC group transformations.
Possible choices of actions for such systems will be considered in Section 4 \footnote{
Different ways of eliminating the auxiliary coset fields by the inverse Higgs effect
and an issue of deriving the relevant constraints as equations of motion from some actions
were discussed in a recent paper~\cite{Arthur}.}.

\subsubsection{Canonical case: coset   ${\mathscr{K}}^{(d)}$ with the stability subalgebra $o(d)$}

We postulate that the constraints of standard conformal mechanics (\ref{IKL-const})
remain valid also in the presence of additional vectorial variables which appear if $d{\neq}0$.
In terms of the covariant derivatives defined by \p{D-all}  the equations (\ref{IKL-const}) take the form
\begin{equation}\label{IKL-const1}
\mbox{(a)}\,\,\,\,\,\mathscr{D}u=0\,,\qquad \qquad \mbox{(b)}\,\,\,\,\,\mathscr{D}z=\gamma^2\,.
\end{equation}
The remaining MC one--forms (\ref{om-A}) are given by the expressions
\begin{eqnarray}
\omega_{A\!,\,i}^+ &=&  d{\mathscr{X}}_{\,i}^{\,+}   \,,\label{om-PF1}\\
\omega_{A\!,\,i}^-&=&  d{\mathscr{X}}_{\,i}^{\,-}  + v_i\,\omega_R^+ =d{\mathscr{X}}_{\,i}^{\,-}  + 2\gamma\,v_i\,\omega_H\,,\label{om-PF2}\\
\omega_{B\!,\,i} &=& dv_i - {\mathscr{X}}_{\,i}^{\,-}\,\omega_R^+ =dv_i  - 2\gamma\,{\mathscr{X}}_{\,i}^{\,-}\,\omega_H\,.\label{om-B1}
\end{eqnarray}
We use here the ``conformal'' basis (\ref{R+-A}), (\ref{Cf-Ab}), since in this case the variable ${\mathscr{X}}_{\,i}^{\,+}$
decouples from other vector variables ${\mathscr{X}}_{\,i}^{\,-}$, $v_i\,$.
It enters only into the one--form $\omega_{A\!,\,i}^+\,$,
whereas the other two MC one--forms contain only  ${\mathscr{X}}_{\,i}^{\,-}$, $v_i\,$.

Besides (\ref{IKL-const}), we also impose the following additional constraints
\begin{equation}\label{add-InvH}
\mbox{(a)}\,\,\,\,\,\omega_{B\!,\,i}= 0\,,\qquad \qquad \mbox{(b)}\,\,\,\,\,\omega_{A\!,\,i}^-=0\,,
\end{equation}
which yield the equations
\begin{equation}\label{eq-y-v}
\mbox{(a)}\,\,\,\,\,\rho^2\dot v_i-2\gamma {\mathscr{X}}_{\,i}^{\,-}=0\,,\qquad
\mbox{(b)}\,\,\,\,\,\rho^2 \dot {\mathscr{X}\,}{}_{i}^{-}+2\gamma v_i=0\,,
\end{equation}
where $\rho=e^{u/2}$.
After eliminating $v_i$ by the inverse Higgs constraint (\ref{eq-y-v}\,b),
we obtain the following new dynamical second-order equations
\begin{equation}\label{eq-X-}
\rho^2\frac{d}{dt}\left(\rho^2\dot{\mathscr{X}\,}{}_{i}^{-}\right) +4\gamma^2 {\mathscr{X}}_{\,i}^{\,-}=0
\end{equation}
for the trajectory functions ${\mathscr{X}}_{\,i}^{\,-}(t)$.

Equations of motions for ${\mathscr{X}}_{\,i}^{\,+}$ are defined by the constraints on the MC one--forms
$\omega_{A\!,\,i}^+$. For instance, the admissible GC covariant constraints are $\omega_{A\!,\,i}^+=0$,
which lead to the constant, time--independent vector ${\mathscr{X}}_{\,i}^{\,+}\,$. It is more interesting to look at the case
when the equations of motion for
${\mathscr{X}}_{\,i}^{\,+}$ are of the second order in time derivative. Such equations are
\begin{equation}\label{eq-X+}
\frac{d}{dt}\left(\rho^2\dot{\mathscr{X}\,}{}_{i}^{+}\right) =0\,.
\end{equation}
These equations, like the equations (\ref{eq-y-v}) and (\ref{eq-X-}), are covariant under the GC transformations (\ref{tr-GC}):
the variations of  (\ref{eq-X+})  are proportional to the equation of motion for the dilaton  (\ref{eq-rho}). For
example,
\begin{equation}\label{var-vec}
\delta \left\{\frac{d}{dt}\left(\rho^2\dot{\mathscr{X}\,}{}_{i}^{+}\right)\right\} =
-\dot\alpha\,\frac{d}{dt}\left(\rho^2\dot{\mathscr{X}\,}{}_{i}^{+}\right)
- \gamma^{-1}\frac{d}{dt}\Big[\rho^3\left(\ddot\rho-\gamma^2\rho^{-3}\right)\delta v_i\Big]\,,
\end{equation}
where $\alpha$ and $\delta v_i$ are defined in (\ref{tr-GC}).

The GC covariance of the equations (\ref{eq-y-v}), (\ref{eq-X-}) and (\ref{eq-X+}) becomes manifest after
rewriting them using the covariant derivatives (\ref{D-new}).
Modulo eqs. (\ref{IKL-const1}), the equations (\ref{eq-y-v}) can be written equivalently as
\begin{equation}\label{eq-y-v1}
\mbox{(a)}\,\,\,\,\,\mathscr{D}v_i=0\,,\qquad \qquad \mbox{(b)}\,\,\,\,\,\mathscr{D}{\mathscr{X}\,}{}_{i}^{-}=0\,,
\end{equation}
while (\ref{eq-X-}) and (\ref{eq-X+}) as
\begin{equation}\label{eq-X-+}
\mbox{(a)}\,\,\,\,\,\mathscr{D}\mathscr{D}{\mathscr{X}\,}{}_{i}^{-}-2\gamma\mathscr{D}v_i=0\,,\qquad \qquad
\mbox{(b)}\,\,\,\,\,\mathscr{D}\mathscr{D}{\mathscr{X}\,}{}_{i}^{+}=0\,,
\end{equation}
where the covariant derivative acting on $\mathscr{D}{\mathscr{X}\,}{}_{i}^{\pm}$ is just $\mathscr{D}=E^{-1}\partial_t$:
$\mathscr{D}\mathscr{D}{\mathscr{X}\,}{}_{i}^{\pm}=\rho^2\partial_t\left(\mathscr{D}{\mathscr{X}\,}{}_{i}^{\pm}\right)\,$.

We see that our extended conformal mechanics is described by the dynamical variables $\rho$ and ${\mathscr{X}}_{\,i}^{\,\pm}$.
The variable $\rho$ still obeys the standard equation (\ref{eq-rho}), but now it is coupled to the vectorial
coset variables ${\mathscr{X}}_{\,i}^{\,\pm}$ via the equations (\ref{eq-X-}) and (\ref{eq-X+}).

There is another dynamical system which is still invariant under the GC symmetry but contains a smaller number of degrees of freedom.
Namely, using the dynamical equations (\ref{eq-X+}) we can consider the system in which, instead of the full
vector ${\mathscr{X}\,}{}_{i}^{-}$, only its covariant projection
\begin{equation}\label{pr-X}
X\equiv \mathscr{X}\,{}_{i}^{-}\,\mathscr{D}{\mathscr{X}\,}{}_{i}^{+}
\end{equation}
appears. Taking into account eqs. (\ref{IKL-const1}), we obtain that
$X{=} \,\rho^2\, \mathscr{X}{}_{i}^{-}\,\dot{\mathscr{X}\,}{}_{i}^{+}$.
The equations (\ref{eq-X-}) lead to the following dynamical equation for $X$:
\begin{equation}\label{eq-X}
\rho^{2}\frac{d}{dt}\left(\rho^2\dot X\right) +4\gamma^2 X=0\,.
\end{equation}
The action for such a system encompassing the dynamical variables $\rho$, ${\mathscr{X}\,}{}_{i}^{+}$ and $X$,
will be presented in Section 4. Note that the equation  (\ref{eq-X}) is the projection of the equations
(\ref{eq-X-+}a) on the covariantly constant vector $\mathscr{D}{\mathscr{X}\,}{}_{i}^{+}$
(see (\ref{eq-X-+}b)). With eqs. (\ref{IKL-const1}) taken into account, the equation (\ref{eq-X})
can be equivalently rewritten in the manifestly covariant form as
\begin{equation}\label{eq-X1}
\mathscr{D}\mathscr{D} X -2\gamma \mathscr{D} V=0\,,
\end{equation}
where $V\equiv v_{i}\,\mathscr{D}{\mathscr{X}\,}{}_{i}^{+}\,$, $\mathscr{D}V = \mathscr{D}v_i\,\mathscr{D}{\mathscr{X}\,}{}_{i}^{+}$
and $\mathscr{D}X =\mathscr{D}{\mathscr{X}\,}{}_{i}^{-}\mathscr{D}{\mathscr{X}\,}{}_{i}^{+}\,$.

\subsubsection{Three non-canonical cosets}

The non-canonical cosets are obtained by including some of the $o(2,1)$ generators $H$, $K$, $D$ into the stability subgroup.
This gives rise to reducing the set of the primary coset fields (\ref{coset-var}).

\bigskip
\noindent $i$) {\sl CGA coset ${\mathscr{K}}^{(d)}_1$ with the stability subalgebra $\left[ o(d)\oplus K \oplus D \right]$.}
\smallskip

This case is obtained by setting $z=u=0$ in the formulae (\ref{om-cm}), (\ref{om-P})-(\ref{om-B}) and  (\ref{coset-var}).
We obtain the following reduced MC one--forms
\begin{equation}\label{om-1}
\omega_H = dt\,,\qquad
\omega_{P\!,\,i} = dx_i - v_i\,dt\,,\qquad
\omega_{F\!,\,i} = df_i \,,\qquad
\omega_{B\!,\,i} = dv_i - 2\,f_i\,dt\,.
\end{equation}
Imposing the inverse Higgs conditions
\begin{equation}\label{inH-1}
\omega_{P\!,\,i} = 0\,,\qquad
\omega_{B\!,\,i} = 0\,,
\end{equation}
we can express $v_i$ and $f_i$ in terms of $x_i$:
\begin{equation}\label{exp-1}
v_i = \dot x_i\,,\qquad
f_i ={\textstyle\frac12}\,\dot v_i ={\textstyle\frac12}\,\ddot x_i\,.
\end{equation}
Further, the additional covariant constraint
\begin{equation}\label{inH-1a}
\omega_{F\!,\,i} = 0
\end{equation}
results in the following dynamical equation for $x_i$
\begin{equation}\label{d-eq-1}
\stackrel{...}{x}_i =0 \,.
\end{equation}
In the $D{=}\,2{+}1$ case, eq. (\ref{d-eq-1}) coincides with the dynamics
of the ``exotic'' model considered in \cite{SZ}, but
here the same equation is obtained for arbitrary $D{=}\,d{+}1$.

\bigskip
\noindent $ii$) {\sl CGA coset ${\mathscr{K}}^{(d)}_2$  with the stability subalgebra $\left[ o(d)\oplus K \right]$.}
\smallskip

This case is obtained by setting $z{=}\,0$ in  (\ref{om-cm}), (\ref{om-P})-(\ref{om-B}) and  (\ref{coset-var}).
The MC one--forms read
\begin{equation}\label{om-cm3}
\omega_H = e^{-u}dt\,,\qquad
\omega_D = du\,,
\end{equation}
\begin{equation}\label{om-P3}
\omega_{P\!,\,i} = e^{-u}\left[ d\left(e^{u}x_i\right) - v_i\,dt \right]\,,\qquad
\omega_{F\!,\,i} = e^{u} d\left(e^{-u}f_i\right) \,,\label{om-F3}\qquad
\omega_{B\!,\,i} = dv_i - 2\,f_i\,e^{-u}dt\,.
\end{equation}
Inverse Higgs conditions $\omega_{P\!,\,i} {=}\, 0$, $\omega_{B\!,\,i} {=}\, 0$ in (\ref{inH-1})
express $v_i$ and $f_i$ in terms of $x_i$ as
\begin{equation}\label{exp-3}
v_i = \left(e^{u}x_i\right)^\cdot \,,\qquad
f_i ={\textstyle\frac12}\,e^{u}\dot v_i\,.
\end{equation}
{}From the condition $\omega_{F\!,\,i} {=} 0$ we get the dynamical equations
\begin{equation}\label{d-eq-3}
\stackrel{...}{y}_i =0
\end{equation}
for $y_i\equiv e^{u} x_i$. Thus, after
the redefinition ${x}_i\to {y}_i\,$, the vector sector coincides with the one
obtained in the case $i$).

Note that we can impose the additional condition $\omega_D {=}\,0$, which implies that the residual variable $u$ becomes a constant.

\bigskip
\noindent $iii$) {\sl CGA coset ${\mathscr{K}}^{(d)}_3$  with the stability subalgebra $\left[ o(d)\oplus D \right]$.}
\smallskip

In this case we set $u{=}\,0$ in the MC one--forms for the canonical coset (\ref{group-el}).
We obtain
\begin{equation}\label{om-cm2}
\omega_H = dt\,,\qquad
\omega_K = dz+z^2 dt\,,
\end{equation}
\begin{eqnarray}
\omega_{P\!,\,i} &=& dx_i -2zx_i\,dt - v_i\,dt\,,\label{om-P2}\\
\omega_{F\!,\,i} &=& df_i +2z f_i\,dt + v_i\,(dz+z^2\,dt)\,,\label{om-F2}\\
\omega_{B\!,\,i} &=& dv_i + 2\,x_i\,(dz+z^2\,dt) - 2\,f_i\,dt\,.\label{om-B2}
\end{eqnarray}
In this case the conditions (\ref{inH-1})
express $v_i$ and $f_i$ in terms of $x_i$ as
\begin{equation}\label{exp-2}
v_i = \dot x_i -2zx_i \,,\qquad
f_i ={\textstyle\frac12}\,\dot v_i +(\dot z +z^2)x_i
\end{equation}
and also lead to the field equations (\ref{d-eq-1}),
which leaves the decoupled variable $z$ arbitrary.
The minimal formulation corresponds to adding the constraint $\omega_K=0$.
In this case we obtain the following dynamical equation for $z$
\begin{equation}\label{d-eq-2z}
\dot z +z^2 =0\,.
\end{equation}

\bigskip
Thus, in all three cases $i$), $ii$) and $iii$), the vector variables
decouple and describe the motion with constant acceleration given by eqs.\,(\ref{d-eq-1}).

\setcounter{equation}{0}
\section{Lagrangean GC--invariant models}

\subsection{New actions for arbitrary $D$}

Here we consider GC invariant actions for arbitrary $D$, without central charge.
We present two GC invariant models. In one of them the Lagrangian is bilinear in the covariant derivatives
of the vector coset variables and the other model is described by the action which resembles
the well-known Brink--Schwarz (BS) action.

\subsubsection{The actions bilinear in covariant derivatives}

We consider the following general class of extended AFF actions
\begin{equation}\label{ac-GI-Y}
S_1= \int dt\, E\, m_{ab}\,\mathscr{D}Y^a_{\,i}\,
\mathscr{D}Y^b_{\,i} \,,
\end{equation}
where $Y^a_{\,i}=(x_i, v_i, f_i)$, $a=1,2,3$ and $m_{ab}$ is a constant matrix. The manifest GC invariance of this action is obvious.
Note that, following the Volkov's proposal \cite{NonLin}, this action can be equivalently rewritten
in geometric way as $\int\frac{m_{ab}\omega^a_i\omega^b_i}{\omega_H}$\,,
 where $\omega^a_i$ are the MC one--forms corresponding
to the variables $Y^a_i$.

Let us study in more detail the example with
\begin{equation}\label{ac-G1}
S_1=\int dt\, L_1= {\textstyle\frac12}\int dt\, E\, \mathscr{D}{\mathscr{X}}_{\,i}^{\,+}
\mathscr{D}{\mathscr{X}}_{\,i}^{\,+}\,,
\end{equation}
where
\begin{equation}\label{D-X+}
\mathscr{D}{\mathscr{X}}_{\,i}^{\,+}=
e^{u} \Big[\dot {\mathscr{X}}_{\,i}^{\,+} - {\mathscr{X}}_{\,i}^{\,-}\left(\dot u -2z \right) +
\gamma^{-1}v_i \Big( e^u\left(\dot z + z^2 \right) -\gamma^2 e^{-u} \Big) \Big]\,,\qquad E=e^{-u}\,.
\end{equation}
One can show that the action (\ref{ac-G1}) describes the dynamical system introduced in Sect.\,3.2.1
from purely geometric considerations. The equations of motion following from  (\ref{ac-G1}) are
\begin{eqnarray}
\delta v_i\,:  \qquad\qquad &  e^u\left(\dot z + z^2 \right) -\gamma^2 e^{-u}=0\,, \label{eq-v}\\
\delta {\mathscr{X}}_{\,i}^{\,-}\,:    \qquad\qquad & \dot u -2z=0\,, \label{eq-y-}\\
\delta {\mathscr{X}}_{\,i}^{\,+}\,:  \qquad\qquad & \dot {\mathscr{P}}_i^+  =0\,,     \label{eq-y+}\\
\delta u\,:  \qquad\qquad &
{\mathscr{P}}_i^+ \left(e^{u}\dot {\mathscr{X}}_{\,i}^{\,-}  +2\gamma v_i\right)
+\frac12\, {\mathscr{P}}_i^+ {\mathscr{P}}_i^+=0\,, \label{eq-u}\\
\delta z\,:  \qquad\qquad &
{\mathscr{P}}_i^+\left(e^{u}\dot v_i -2\gamma {\mathscr{X}}_{\,i}^{\,-}\right)=0\,,  \label{eq-z}
\end{eqnarray}
for $\mathscr{D}{\mathscr{X}}_{\,i}^{\,+}\neq 0$. Here
\begin{equation}\label{mom-X+a}
{\mathscr{P}}_i^+ = \mathscr{D}{\mathscr{X}}_{\,i}^{\,+}
\end{equation}
is the momentum conjugate to $ {\mathscr{X}}_{\,i}^{\,+}\,$.
Eqs. (\ref{eq-v}) and  (\ref{eq-y-}) provide the relation
\begin{equation}\label{mom-X+}
{\mathscr{P}}_i^+ =  e^{u} \dot {\mathscr{X}}_{\,i}^{\,+}\,.
\end{equation}

The equations  (\ref{eq-v}) and  (\ref{eq-y-}) are the equations of standard conformal mechanics (see (\ref{z-rez})
and (\ref{eq-rho})).
Thus, the action (\ref{ac-G1})
reproduces as well the equations of motion for the standard conformal mechanics sector.
Eliminating $v_i$ from the equations  (\ref{eq-u}) and (\ref{eq-z}), we obtain
\begin{equation}\label{pr-eq-X-}
{\mathscr{P}}_i^+ \left[\frac{d}{dt}\left(\rho^2\dot{\mathscr{X}\,}{}_{i}^{-}\right) +4\gamma^2\rho^{-2}{\mathscr{X}}_{\,i}^{\,-}\right]=0\,.
\end{equation}
This is the projection
of the field equations (\ref{eq-X-}) for ${\mathscr{X}}_{\,i}^{\,-}$ on ${\mathscr{P}}_{\,i}^{\,+}$,
i.e. we obtained precisely eq.~(\ref{eq-X}).
Finally, eqs. (\ref{eq-y+}) are the equations of motion (\ref{eq-X+}) for ${\mathscr{X}}_{\,i}^{\,+}$.

Thus the model with the action \p{ac-G1} amounts to one
of the dynamical systems described in Sect.\,3.2.1 by constrained MC one--forms.
This particular system is represented by the geometric variables $\rho$, ${\mathscr{X}}_{\,i}^{\,+}$ and $X$
with the equations of motion  (\ref{eq-rho}), (\ref{eq-X+}) and (\ref{eq-X}).
At present it is not known whether one can define the off-shell action for the system in which all
${\mathscr{X}}_{\,i}^{\,-}$ are dynamical and are described by the equations (\ref{eq-X-}).

Eliminating auxiliary variable $z$ by the algebraic equation (\ref{eq-y-})
and introducing the new variable
$
\rho=e^{u/2}\,,
$
we obtain an equivalent Lagrangian
\begin{equation}\label{Lagr-comp1a}
L_1 =  {\textstyle\frac12} \,\rho^{2} \Big[\dot{\mathscr{X}}_{\,i}^{\,+} +
\gamma^{-1}v_i \Big( \rho\ddot\rho -\gamma^2 \rho^{-2} \Big) \Big]^2.
\end{equation}
The equations of motion following from (\ref{Lagr-comp1a}) can be identified with those derived from (\ref{ac-G1}).

Let us now consider the Hamiltonian formulation of the system described by the action (\ref{ac-G1}).
The definitions of the momenta lead to the primary constraints
\begin{eqnarray}
& {\mathscr{P}}_{v_i}   \approx  0\,,\qquad\qquad\qquad\qquad & {\mathscr{P}}_i^-   \approx  0\,,\qquad\label{const-v}\\
& {\mathscr{F}}_u  \equiv  p_u +{\mathscr{X}}_i^- {\mathscr{P}}_i^+  \approx 0\,, \qquad
& {\mathscr{F}}_z  \equiv  p_z-\gamma^{-1}e^u v_i {\mathscr{P}}_i^+  \approx  0\,,\label{const-u}
\end{eqnarray}
where the canonical pairs are
\begin{equation}\label{cPB}
\{ {\mathscr{X}}_i^\pm, {\mathscr{P}}_j^\pm\}_{{}_P}=\delta_{ij}\,,\qquad
\{ v_i, {\mathscr{P}}_{v_j}\}_{{}_P}=\delta_{ij}\,,\qquad
\{ z, p_z\}_{{}_P}=1\,,\qquad
\{ u, p_u\}_{{}_P}=1\,.
\end{equation}
The canonical Hamiltonian is
\begin{equation}\label{canHam}
H_1 =  {\textstyle\frac12}\,e^{-u}  {\mathscr{P}}_i^+ {\mathscr{P}}_i^+
-\left[2z {\mathscr{X}}_i^- + \gamma^{-1}v_i \Big( e^u z^2  -\gamma^2 e^{-u} \Big)\right]  {\mathscr{P}}_i^+\,. \nonumber
\end{equation}

{}From the explicit form of non-vanishing PB of the constraints (\ref{const-v})--(\ref{const-u})
\begin{equation}\label{nonPB-const}
\{ {\mathscr{F}}_u, {\mathscr{P}}_i^-\}_{{}_P}={\mathscr{P}}_i^+\,,\qquad
\{ {\mathscr{F}}_z, {\mathscr{P}}_{v_i}\}_{{}_P}=-\gamma^{-1}e^u{\mathscr{P}}_i^+\,,\qquad
\{ {\mathscr{F}}_u, {\mathscr{F}}_{z}\}_{{}_P}=\gamma^{-1}e^uv_i{\mathscr{P}}_i^+
\end{equation}
we  see that  the constraints (\ref{const-v})--(\ref{const-u})
are the mixture of first and second class constraints.
A simple analysis shows that
the considered system is described by the second class constraints ${\mathscr{F}}_u\approx0$, ${\mathscr{F}}_z\approx0$,
${\mathscr{P}}_{v_i}{\mathscr{P}}_i^+\approx0$, ${\mathscr{P}}_i^-{\mathscr{P}}_i^+\approx0$
and the first class constraints given by the  components of
${\mathscr{P}}_{v_i}\approx0$ and ${\mathscr{P}}_i^-\approx0$ orthogonal to ${\mathscr{P}}_i^+$.
Using the gauge freedom generated by the first class constraints we can
eliminate the components of
$v_{i}$, ${\mathscr{P}}_{v_i}$,  ${\mathscr{X}}_i^-$ and ${\mathscr{P}}_i^-$ orthogonal to ${\mathscr{P}}_i^+$.

Remaining phase space variables are  $\tilde{\mathscr{X}}_i^+$, ${\mathscr{P}}_i^+$, $u$, $p_u$, $z$, $p_z$
and the projections
\begin{equation}\label{new-sc}
V\equiv v_i {\mathscr{P}}_i^+\,,\quad P_V\,;\qquad
X\equiv {\mathscr{X}}_i^- {\mathscr{P}}_i^+\,,\quad P_X
\end{equation}
of $v_i$, ${\mathscr{P}}_{v_i}$, $ {\mathscr{X}}_i^-$, ${\mathscr{P}}_i^-$.
The expressions for the new variables $\tilde{\mathscr{X}}_i^+$, $P_V$ and $P_X$ can be given explicitly.
It is important to note that, if we introduce Dirac brackets (DB),
for the remaining variables they coincide  with canonical Poisson brackets.
The remaining second class constraints take the form
\begin{equation}\label{res-const}
P_V  \approx  0\,, \qquad
P_X  \approx 0\,, \qquad\qquad
{\mathscr{F}}_u  \equiv  p_u +X  \approx 0\,, \qquad
{\mathscr{F}}_z  \equiv p_z-\gamma^{-1}e^u V  \approx  0\,.
\end{equation}

Introducing DB for the second class constraints (\ref{res-const})
and eliminating the variables $p_u$, $p_z$, $P_V$ and $P_X$,
we are left with the variables $u$, $z$, $V$ and $X$ with the following non-vanishing DB
\begin{equation}\label{DB-rem-var}
\{ u, X \}_{{}_D}=-1 \,,\quad \{ z, V \}_{{}_D}=\gamma e^{-u} \,,\quad
\{ V, X \}_{{}_D}= V\,, \quad \{ {\mathscr{X}}_i^+, {\mathscr{P}}_j^+ \}_{{}_D}=\delta_{ij}
\end{equation}
and the Hamiltonian
\begin{equation}\label{tHam-res}
H_1 = {\textstyle\frac12}\,e^{-u}  {\mathscr{P}}_i^+ {\mathscr{P}}_i^+
-2z X -  \gamma^{-1} \Big( e^u z^2  -\gamma^2 e^{-u} \Big)V \,.
\end{equation}
The set of equations (\ref{DB-rem-var}), (\ref{tHam-res}) determines the dynamics of our model (\ref{ac-G1}) in phase space.
We can check that the equations of motion generated by the Hamiltonian (\ref{tHam-res}),
$\dot u=\{ u, H \}_{{}_D}\,$, etc, coincide with eqs. (\ref{eq-v})-(\ref{eq-z}).

Introducing the variables
\begin{equation}\label{new-sc1}
\rho\equiv e^{u/2}\,,\qquad p_\rho \equiv -2e^{-u/2}X,\qquad
y \equiv  -2e^{u}V \,,\qquad p_y\equiv \gamma^{-1}z,
\end{equation}
which form two canonical pairs,
\begin{equation}\label{DB-rem-var1}
\{ \rho, p_\rho \}_{{}_D}=1 \,,\qquad \{ y, p_y \}_{{}_D}=1 \,,
\end{equation}
(other DBs are vanishing) we can put the Hamiltonian (\ref{tHam-res}) in the following form
\begin{equation}\label{tHam-res1}
H_1= {\textstyle\frac12}\,\rho^{-2}  {\mathscr{P}}_i^+ {\mathscr{P}}_i^+
+\gamma \left( \rho p_\rho+y p_y\right) p_y -\gamma \rho^{-4}y
\,.
\end{equation}
Now we can present our model in a more economical formulation. Namely, we can use the following first-order Hamiltonian
form of the action:
\begin{equation}\label{1-order}
S_1=\int dt \left[ {\mathscr{P}}_i^+  \dot{\mathscr{X}}_i^+ +
p_\rho \dot \rho +
p_y \dot y -{\textstyle\frac12}\,\rho^{-2}  {\mathscr{P}}_i^+ {\mathscr{P}}_i^+
-\gamma \left( \rho p_\rho+y p_y\right) p_y +\gamma \rho^{-4}y\right].
\end{equation}
Eliminating momenta ${\mathscr{P}}_i^+$, $p_\rho$ and $p_y$ by their equations of motion, we finally obtain
\begin{equation}\label{2-order}
S_1=\int dt \left[\, \frac12\,\rho^2 \dot{\mathscr{X}}_i^+ \dot{\mathscr{X}}_i^+ +
\frac{1}{\gamma\rho}\left(\dot y\dot\rho-\frac{y}{\rho}\, \dot\rho\dot\rho\right) +
\frac{\gamma y}{\rho^4}\,\right].
\end{equation}
This new action is a generalization of the conformal mechanics action (\ref{ac-AFF}).
Besides invariance under the one--dimensional conformal symmetry ${\rm SO}(2,1)$ acting on $\rho$ and $y$
(recall the definitions (\ref{new-sc1})), the model  (\ref{2-order}) is invariant under
the full GC symmetry with $d{\neq}\,0$.

\subsubsection{Square root action}

The second way of introducing the dynamics in the sector of vector coset parameters provides
the Lagrangian as the square root of
the product of vector one--forms, in a way resembling the model of free relativistic particle. We consider the action
\begin{equation}\label{ac-G2}
S_2= m \int \sqrt{\omega_{A\!,\,i}^+ \omega_{A\!,\,i}^+}\;,
\end{equation}
where $m$ is a constant. It is a particular case of more general action
\begin{equation}\label{ac-GI-Y2}
S_2=m\int \sqrt{ m_{ab}\,\omega^a_i\omega^b_i} \;,
\end{equation}
where $\omega^a_i$ are vector MC one--forms.

By applying variational principle to  the action (\ref{ac-G2}) we get eqs. (\ref{eq-v})--(\ref{eq-z})
with the minor modification in eq.~(\ref{eq-u}),
\begin{equation}\label{eq-u2}
\delta u\,:  \qquad\qquad
{\mathscr{P}}_i^+ \left(e^{u}\dot {\mathscr{X}}_{\,i}^{\,-}  +2\gamma v_i\right)=0\,,
\end{equation}
and with the additional property that in all equations the expression
\begin{equation}\label{mom-X+2}
{\mathscr{P}}_i^+ =  m\,\mathscr{D}{\mathscr{X}\,}{}_{i}^{+}
\Big(\mathscr{D}{\mathscr{X}\,}{}_{k}^{+}\mathscr{D}{\mathscr{X}\,}{}_{k}^{+}\Big)^{-1/2}
\end{equation}
should be substituted in place of the variable ${\mathscr{P}}_i^+\,$.
As in the previous model, from eqs.~(\ref{eq-v}), (\ref{eq-y-}) follows that
$\mathscr{D}{\mathscr{X}\,}{}_{i}^{+} =  e^{u}\dot{\mathscr{X}}_{\,i}^{\,+}\,$,
and as well ${\mathscr{P}}_i^+ =  m\,\dot{\mathscr{X}}_{\,i}^{\,+}
\left(\dot{\mathscr{X}}_{\,k}^{\,+}\dot{\mathscr{X}}_{\,k}^{\,+}\right)^{-1/2}\!$.
{}From the latter expression for ${\mathscr{P}}_i^+ $
we derive the important relation
\begin{equation}\label{const-P2}
{\mathscr{M}}  \equiv  {\mathscr{P}}_i^+ {\mathscr{P}}_i^+ -m^2  =0\,.
\end{equation}

Despite the difference between the equations (\ref{eq-u2}) and
(\ref{eq-u}), after elimination of $v_i$ eqs.~(\ref{eq-u2}) and
(\ref{eq-z}) yield the same equation (\ref{pr-eq-X-}) for $X{=}
{\mathscr{X}}_i^- {\mathscr{P}}_i^+$
 \footnote{Note that adding of the ``cosmological'' term
$(-\frac12 \, m^2\! \int e^{-u})$ to the action (\ref{ac-G2}) leads
to the new term $(\frac12 \, m^2)$ in the l.h.s. of (\ref{eq-u2}) which coincides, due to (\ref{const-P2}),
with the term $(\frac12 \, {\mathscr{P}}_i^+{\mathscr{P}}_i^+)$ in (\ref{eq-u}). }.
Thus the two--parameter sector $(\rho,X)$ is described by the same equations as in  Sect. 4.1.1:
by the AFF equation  (\ref{eq-rho}) and the equation  (\ref{pr-eq-X-}).

However, the dynamics in the sector of vector variable ${\mathscr{X}}_{\,i}^{\,+}$ is now different.
Distinctly from the model considered in Sect. 4.1.1,
the quantities (\ref{mom-X+2}),
which become the conjugate momenta for $\mathscr{X}_{\,k}^{\,+}$ in the Hamiltonian formalism,
are constrained by eq. \p{const-P2}. As a consequence of this constraint,
and taking into account eqs. (\ref{eq-v}) and (\ref{eq-y-}),
the equations (\ref{eq-y+}) prove to be linearly dependent
\begin{equation}\label{Not-id}
\dot{\mathscr{X}}_{\,i}^{\,+} \dot {\mathscr{P}}_i^+\equiv 0\,.
\end{equation}
The condition (\ref{const-P2}) becomes transparent in the Hamiltonian language,
where it appears as the additional  first class constraint.

The definitions of the momenta yield the same set of primary constraints (\ref{const-v}), (\ref{const-u}) and
also the additional constraint ${\mathscr{M}} \approx 0$ (\ref{const-P2}).
The last constraint indicates that momentum vector ${\mathscr{P}}_i^+$ parametrizes
the sphere $S^{\,d-1}$.
The constant $m$ plays the role of its radius (in the case of BS superparticle described by a similar square-root action
an analogous constant is identified with a mass of the relativistic particle).
The canonical Hamiltonian of the square-root system (\ref{ac-G2}) is given by the second term in (\ref{canHam})
\begin{equation} \label{canHam2}
H_2 =
-\left[2z {\mathscr{X}}_i^- + \gamma^{-1}v_i \Big( e^u z^2  -\gamma^2 e^{-u} \Big)\right]  {\mathscr{P}}_i^+\,.
\end{equation}

The constraint (\ref{const-P2}) has the vanishing PB with all other constraints (\ref{const-v}), (\ref{const-u}).
Thus, the set of constraints is the same as in the model (\ref{ac-G1}). The only difference between these two systems is the presence
of additional first class ``mass'' constraint (\ref{const-P2}) in the second system.
Similarly to the  previous case, after gauge-fixing for the first class constraints presented in (\ref{const-v}), (\ref{const-u}), eliminating
the auxiliary phase space variables and introducing new variables (\ref{new-sc1}) which are canonical with respect to DB,
we find that the ``square-root'' model (\ref{ac-G2}) is described by the following Hamiltonian
(compare with (\ref{tHam-res1})):
\begin{equation}\label{tHam-res2}
H_2= \lambda\left( {\mathscr{P}}_i^+ {\mathscr{P}}_i^+ -m^2\right) +\gamma \left( \rho p_\rho+y p_y\right) p_y -\gamma \rho^{-4}y\,.
\end{equation}
Here, the first term reflects the presence of the first class constraint (\ref{const-P2}), with $\lambda(t)$ being a Lagrange multiplier.

\bigskip

To summarize, in Sect.\,4.1 we illustrated the method of nonlinear realizations on the simplest particular cases of the general actions (\ref{ac-G1})
and (\ref{ac-G2}). The study of these general actions, and, perhaps, of their further extensions,
with added actions for the scalar
Goldstone fields $u$, $z\,$, deserves further studies. It is important to note that the additional
vector variables ${\mathscr{X}}_i^+$ and ${\mathscr{P}}_i^+\,$, which are the characteristic feature of the new GCA
invariant  models,  can be presumably treated as a kind of `angular' variables, keeping in mind that, in
multidimensional mechanical models with ${\rm SO}(2,1)$ invariance,
the standard conformal mechanics describes a radial variable.

\subsection{``Exotic''  $D{=}\,2{+}1$ case}

In this case, the presence of the MC one--form associated with the central charge makes it possible to consider
GC mechanics described by the action
\begin{equation}\label{act-eG}
\tilde {S\,\,}=\tilde {S}_{conf}+\tilde {{S}_\theta}=-\gamma\int \omega_{R}^+ +\theta\int\omega_{\Theta} \,.
\end{equation}
The first term defines the standard conformal mechanics sector \cite{AFF,IKL}, whereas the sector of the vector coset fields
is represented by the WZ term.

To obtain a minimal formulation we use the inverse Higgs conditions
\begin{equation}\label{iHC-eG}
\omega_{P\!,\,i} = 0\,,\qquad \omega_{B\!,\,i} = 0
\end{equation}
in the action (\ref{act-eG}).

First we consider the case with the stability subalgebra $\left[ o(d)\oplus K \oplus D \right]$ (case $i$)).
The solutions of the constraints (\ref{iHC-eG}) are given in (\ref{exp-1}).
Inserting these solutions in the action (\ref{act-eG}), we obtain, modulo a total derivative under the integral,  the action
\begin{equation}\label{ac-eG-con}
\tilde S_{\theta}= {\textstyle\frac{1}{2}}\,\theta\!\int dt\,\epsilon_{ij}\dot x_i \ddot{x}_j \,.
\end{equation}
It is the action of the ``massless'' particle with the higher-order
Chern--Simons--type term \cite{SZ}, which is the $m\to 0$ limit  of the action  (\ref{ac-eG-c}).
Other choices $ii$), $iii$) of the stability subalgebra, i.e. $\left[ o(d)\oplus K  \right]$ or
$\left[ o(d) \oplus D \right]$, lead to the same result: the dynamics is again described
by the action (\ref{ac-eG-con}) [in the case $ii$) -- after redefining the variable $x_i$].

In the canonical case, when the stability subalgebra does not contain the $o(2,1)$ generators, the first term in  (\ref{act-eG})
produces dynamics in the $o(2,1)$ sector. Inserting in the action (\ref{act-eG}) the expressions
\begin{equation}\label{exp-eL}
v_i = e^u\left[\dot x_i +(\dot u-2z)x_i \right] \,,\qquad
f_i =e^u\left[{\textstyle\frac12}\,\dot v_i +e^u(\dot z +z^2)x_i\right],
\end{equation}
which follow from the inverse Higgs constraints (\ref{iHC-eG}),
as well as the expressions for the one--forms (\ref{om-cm}), we obtain
\begin{equation}\label{exp-w-th}
\omega_\Theta = \left[ {\textstyle\frac{1}{2}}\,\epsilon_{ij}\dot y_i \ddot{y}_j +
\frac{d}{dt}\left(\phi-z\epsilon_{ij}y_i \dot{y}_j \right) \right] dt \,.
\end{equation}
Then, eliminating the field $z$ by its algebraic equation of motion,
we obtain (modulo a total derivative) the action
\begin{equation}\label{ac-eG-con-gen}
\tilde S=\int dt \left( \dot\rho^2-\frac{\gamma^2}{\rho^2} +
\frac{\theta}{2}\,\epsilon_{ij}\dot y_i \ddot{y}_j \right),
\end{equation}
where $y_i\equiv e^u x_i$\,.

Thus we ended up with a decoupled pair of the GC--invariant  $D{=}\,2{+}1$ models. One of them is the AFF conformal mechanics
with the action (\ref{ac-AFF}) \cite{AFF}, and the other one is described by the WZ action (\ref{ac-eG-con}),
firstly proposed in \cite{SZ}.

\setcounter{equation}{0}
\section{Conclusions}

We have investigated nonlinear realizations of the Galilean conformal group in arbitrary
space--time dimensions $D$, including the ``exotic'' $D{=}\,2{+}1$ case with the additional central charge.
The analysis of the MC one--forms with the appropriate inverse Higgs and dynamical covariant constraints
in many cases is sufficient to reveal the underlying dynamics of the new mechanical systems
with Galilean conformal symmetry. Alternatively, one can use the MC one--forms
for the construction of the invariant actions and obtaining some further examples of the Galilean conformal mechanics.
Following \cite{AIM-F,GGT,AGKP}, in the $D{=}\,2{+}1$ case we use the central charge MC one--form to describe the WZ term in the action.

We recall that, until recently, the only known GC--invariant classical mechanics Lagrangian model was the
one related to the ``exotic'' dimension $D{=}\,2{+}1$, that is the non-relativistic particle model described
by the action with the higher order Chern--Simons--type term \cite{LSZ,SZ} \footnote{
For completeness, it should be mentioned that the field equations of magnetic--like Galilean
electrodynamics \cite{LeBel} are also covariant under GCA  \cite{DuHo}.
Furthermore, one can construct the systems of nonlinear
partial differential equations invariant under GCA (as was shown in \cite{Cher}
for any $d$ and in \cite{Cher,MarTach} for the planar $d{=}2$ case, with GCA enlarged by ``exotic''
central charge).}.
The powerful techniques of the nonlinear realizations allowed us to obtain the whole family
of new models exhibiting Galilean conformal symmetry for any  $D{=}\,d{+}1$.
Moreover, using the covariant MC one--forms, it is possible to
construct many such models -- we considered only some simple examples.

The difficulty with finding dynamical realizations of GCA with $d{\neq}\,0$ was one of the undesirable features
of this symmetry. In this paper, by using the nonlinear realizations approach,
we obtained new dynamical realizations, including those for an abitrary space dimension $d$.
We proposed three GC--invariant models of classical mechanics which contain, besides
the scalar coset coordinate $\rho$, also  non-relativistic vector coordinates.
In Sect.\,4 we obtained the known model of AFF conformal mechanics
(in the first-order formalism, with the original degrees of freedom $u=\frac12 \ln\rho$ and $z\,$), accompanied
by couplings to the additional non-relativistic vector variables ${\mathscr{X}}_{\,i}^{\,\pm}\,$.
It is interesting that in this case one can define two actions with different
Lagrangians which lead to similar dynamical equations,
such that in both cases the sector of the  AFF conformal mechanics is decoupled.
One more model (Sect.\,4.2) is specific for the $D{=}\,2{+}1$ case. It involves the GC--covariant coupling
between the degrees of freedom $u$, $z$ and the non-relativistic vector coordinate $x_i$ through the WZ term defined
by the MC one-form associated with the ``exotic'' central charge.
It turns out that  the conformal mechanics degrees of freedom $(u,z)$ and
the vectorial ones $y_i=e^u x_i$ decouple again in this model.
We would like to add that the model (\ref{ac-eG-con-gen}) in $D{=}\,2{+}1$ contains
higher (third) order time derivative, while the field equations (\ref{eq-v})--(\ref{eq-z})
of the first two models are of the first and second orders only.

In this paper we studied the Galilean conformally
invariant models at the classical level, based on the geometric
properties of the Galilean conformal symmetry. The next step
will consist in analyzing quantum properties of the new mechanical
systems constructed in our paper.
In particular, the quantized version of simple model
(\ref{ac-eG-c}) in $D\,{=}\,2{+}1$ space--time was studied earlier \cite{LSZ,SZ}.
As was shown in \cite{LSZ}, despite the presence of the states
with negative norm in this model due to higher-order time derivatives in the action,
it is possible to remove such states by imposing the appropriate
constraints and maintain unitarity in physical subspace of states. As a mechanical
model on the non-commutative
two-dimensional plane, the model (\ref{ac-eG-c}) reveals also direct links to the
description of anyons, to the quantum Hall effect and related issues of
the condensed-matter physics (see, e.g.,
\cite{HorPlV,HorMarSt} and references therein). In
the subsequent studies, we plan to elaborate in similar contexts on the quantum properties
of the new models presented here and to consider as well their field--theoretical extensions.

Finally we add that we did not address in this paper
supersymmetric generalizations of the Galilean conformal algebra \cite{AL,Sa} which
should yield extensions of the $D{=}\,0{+}1$ superconformal mechanics models. Currently, such extensions are under consideration.
Also, it would be interesting to perform the quantization, to find the quantum spectrum
of the new GC--invariant models and to clarify the role of the additional vector variables in physical considerations.

\section*{Acknowledgements}

\noindent We acknowledge a support from a grant of the Bogoliubov-Infeld Programme and RFBR
grants 09-02-01209, 09-01-93107 (S.F. \&\ E.I.), as well as
from the Polish Ministry of Science and Higher Educations grant No.~N202331139 (J.L.).
S.F. thanks the Institute of Theoretical Physics at the Wroclaw University
for the warm hospitality extended to him during the course of this study.

\end{document}